\documentclass[aps,pre,reprint,superscriptaddress,nofootinbib]{revtex4-2}

\usepackage{amsmath,amssymb,bm,mathtools}
\usepackage{graphicx}
\usepackage{hyperref}
\hypersetup{
  colorlinks=true,
  linkcolor=blue,
  citecolor=blue,
  urlcolor=blue,
  pdftitle={A microscopic heat engine with many hidden variables: stall dissipation, observable inference, and multidimensional bounds},
  pdfauthor={Mesfin Asfaw Taye}
}

\newcommand{\ep}{\dot e_{\mathrm{p}}}
\newcommand{\st}{\mathrm{st}}
\newcommand{\sstate}{\mathrm{ss}}
\newcommand{\hid}{\mathrm{hid}}
\newcommand{\vis}{\mathrm{vis}}
\newcommand{\obs}{\mathrm{obs}}
\newcommand{\Var}{\mathrm{Var}}
\newcommand{\dd}{\mathrm{d}}
\newcommand{\avg}[1]{\left\langle #1 \right\rangle}

\begin{document}

\title{A microscopic heat engine with many hidden variables: stall dissipation, observable inference, and multidimensional bounds}

\author{Mesfin Asfaw Taye}
\affiliation{Science Division, West Los Angeles College, Culver City,
California 90230, USA}

\begin{abstract}
A microscopic motor is usually monitored through a single mechanical
coordinate, while the chemical, conformational, or rotational cycles that power
it remain unresolved. When the measured velocity vanishes the motor is said to
stall, but mechanical silence need not imply thermodynamic equilibrium. We
study an overdamped motor in which one observed coordinate is coupled
reciprocally to several hidden phases through a shared periodic interaction. The
rank-one force--torque structure yields an exact pointwise identity between the
observed and hidden probability currents and an exact decomposition of the
stall entropy production into a reciprocal part and an orthogonal hidden part.
The reciprocal contribution is fixed by a current-square functional
$\Phi_x^{\mathrm{st}}$, the stall load, and the calibrated coupling, whereas
hidden currents orthogonal to the coupling direction remain irreducible. We
then remove an important observational limitation of the static formulation:
for the constant-mobility reciprocal model, the heat dissipated through the
observed bath is exactly $T\Phi_x$ at every load. The Harada--Sasa relation
therefore reconstructs $\Phi_x$ from the velocity correlation and linear
response of $x(t)$ alone, while the time-reversal asymmetry of an unperturbed
$x$ trajectory provides a passive lower bound on the total dissipation. We also
generalize the current identity, orthogonal decomposition, and lower bound to
several observed interaction coordinates coupled by a matrix. The dimension of
the invisible hidden-current space is the nullity of that coupling matrix.
Configuration-dependent mobilities preserve a local identity but introduce an
explicit mobility--current covariance that prevents the constant-mobility
closure without additional information. Direct Langevin simulations of a
reciprocally geared two-phase motor verify the exact stall load, the persistence
of local observed currents at zero mean velocity, and the quadratic increase of
dissipation caused by an orthogonal hidden circulation that leaves all observed
statistics unchanged. An exactly solvable nonreciprocal clock and a
two-reservoir heat-engine bookkeeping complete the comparison and show that
positive hidden dissipation lowers the efficiency below Carnot. The results
separate static, passive-trajectory, and active-response levels of inference and
state precisely what a stalled observed coordinate can and cannot reveal about
a nonequilibrium machine.
\end{abstract}

\maketitle

\section{Introduction}
\label{sec:intro}

Small engines operate in a regime where thermal fluctuations, mechanical load,
and internal chemical or conformational transitions are all of comparable
importance, so that their behavior must be described by probability currents
and entropy production rather than by deterministic force balance
\cite{Seifert2012,Sekimoto2010}. Directed transport in periodic and ratchet
potentials, and the conversion of thermal or chemical free energy into work by
Brownian motors, have been analyzed extensively both as models of molecular
machines and as autonomous heat engines
\cite{Reimann2002,Hanggi2009,Asfaw2002AdjustableBrownianHeat,Asfaw2004CurrentMaximumPower,Asfaw2005EnergeticsSimpleMicroscopic,Asfaw2007ExploringOperationTiny,Asfaw2008ModelingEfficientBrownian,Taye2021BrownianMotorsArranged}.
Exact expressions for the current, the efficiency, and the entropy production
have been obtained for a range of such systems, including engines with
spatially varying temperature and time-dependent driving
\cite{Asfaw2014ThermodynamicFeatureBrownian,Taye2017IrreversibleBrownianHeat,Taye2022ExactTimeDependent2,Taye2024ExactTimeDependent,Taye2015EffectTemperatureDependence,Duki2015EffectTemperatureViscous,Taye2023TimeDependentSolutions,Taye2025ThermodynamicFeaturesHeat},
and the finite-time and finite-power operation of such devices connects
naturally to endoreversible efficiency bounds
\cite{CurzonAhlborn1975,Taye2025CurzonAhlbornType}. The associated entropy
balance, relating the entropy production rate to the entropy extracted into the
reservoirs, has been derived in closed form for overdamped and underdamped
dynamics with static and driven potentials
\cite{Taye2016FreeEnergyEntropy,Taye2015ExactAnalyticalThermodynamic,Taye2015ExactAnalyticalExpressions,Taye2020EntropyProductionEntropy,Taye2022ExactTimeDependent,Taye2023TimeDependentThermodynamic,Taye2024ExactTimeDependent2,taye2025EntropyProductionThermodynamic,Taye2026EntropyProductionMacroscopic,Taye2025ThermodynamicIrreversibilityUnderdamped,Taye2021EffectViscousFriction}.

A feature common to nearly all experimental realizations is partial
observation. The measured quantity is frequently a single coarse coordinate:
the displacement of a probe bead, the step of a motor along a track, or a
reaction coordinate, while the internal state that supplies the driving is not
resolved. Coarse-graining a nonequilibrium system over such hidden degrees of
freedom is known to modify the apparent thermodynamics, and the entropy
produced along hidden transitions is generally not captured by the observed
coordinate alone \cite{Esposito2010,HorowitzEsposito2014,Schnakenberg1976}.
Complementary information about the hidden activity can sometimes be recovered
from the breakdown of the fluctuation--response relation in the observed
channel \cite{Harada2005}, and the interplay of active and passive internal
modes, of internal coordinates coupled to the observed one, and of hidden
switching between operating modes has been examined in several specific motor
models
\cite{Taye2025CompetingActivePassive,Taye2025EntropyProductionThermodynamic,Taye2026ExactThermodynamicAnalysis,Asfaw2014ThermallyActivatedBarrier,Asfaw2012ExploringDynamicsDimer,Asfaw2011NoiseCreatedBistability,Taye2026NoiseActivatedDopant}.

Partial observation is most consequential at mechanical stall, the operating
point at which the measured current vanishes and the motor delivers no output
power. In a fully observed one-dimensional periodic system, a vanishing steady
current forces the steady probability current to vanish everywhere and stall
coincides with equilibrium. This coincidence does not survive coarse-graining.
Because the entropy production rate is a non-negative quadratic functional of
the full probability-current field, a single scalar condition on a projected
current cannot force the vector current field to vanish. The observed coordinate
can therefore be silent while the full system circulates. What has been missing
is a controlled statement of how much of the hidden dissipation at stall is
fixed by the observed coordinate and how much is irretrievably hidden, together
with the structural condition that separates the two. General bounds relating
current, dynamical activity, and entropy production
\cite{Taye2026UniversalThermodynamicInequality,Taye2025UnifiedNonequilibriumFramework}
constrain the observable channel but do not, by themselves, reconstruct hidden
currents; and generic hidden variables need not couple back to the observed
coordinate at all.

This paper addresses that gap and then goes beyond the static reconstruction.
We first study an overdamped motor in which one observed coordinate is coupled
to $N$ internal phases through a single shared interaction potential. The same
energy generates the force on the observed coordinate and the reaction torques
on the phases, producing an exact pointwise reciprocal-current identity. From
this identity we obtain the full entropy production at stall, a weighted
parallel--orthogonal decomposition of the hidden currents, and a sharp lower
bound. The orthogonal component is a genuine no-go sector: it can dissipate
without changing any dynamics in the observed coupling direction.

The stationary one-point marginal and the mean displacement current do not
determine the reciprocal current-square functional $\Phi_x$. Nevertheless, the
full time-dependent observed coordinate contains more information. We prove
that, for the constant-mobility reciprocal model, $T\Phi_x$ is exactly the heat
dissipated through the observed bath. The Harada--Sasa equality
\cite{Harada2005} then converts $\Phi_x$ into an $x$-only fluctuation--response
measurement. Without an applied probe, the Kullback--Leibler rate between the
forward and time-reversed observed trajectories gives a rigorous lower bound on
total entropy production \cite{Gaspard2004,RoldanParrondo2010}, although it need
not detect an orthogonal hidden circulation.

We next replace the scalar gearing lengths by a coupling matrix between several
observed interaction coordinates and many hidden phases. The local current
identity survives in matrix form. A mobility-weighted pseudoinverse gives the
reconstructable hidden component, the null space gives the invisible component,
and a Cauchy--Schwarz argument yields a multidimensional stall-dissipation
bound. We further state the exact correction generated by
configuration-dependent mobilities: the pointwise relation remains valid, but
stall no longer removes a mobility--current covariance, so additional calibrated
information is required. Finally, we test the reciprocal theory directly by
simulating a two-phase geared motor. The simulation separates a fixed observed
stall signature from a tunable orthogonal hidden current and confirms the
predicted quadratic excess dissipation. For comparison, we retain an exactly
solvable nonreciprocal clock and the two-reservoir heat-engine bookkeeping,
which expose the same distinction between mechanical stall and reversibility
without the reciprocal closure.

The remainder of the paper is organized as follows.
Section~\ref{sec:model} defines the reciprocal multi-phase model and its
Langevin dynamics. Section~\ref{sec:framework} gives the Fokker--Planck,
entropy, and power balances. Section~\ref{sec:reduction} reduces the dynamics to
the coupling coordinate and obtains the stall condition.
Section~\ref{sec:local} derives the local current identity and exact stall
entropy production, and Sec.~\ref{sec:decomp} gives the
parallel--orthogonal decomposition and lower bounds.
Section~\ref{sec:solvable} treats the nonreciprocal clock and
Sec.~\ref{sec:heat} gives the heat-engine bookkeeping.
Section~\ref{sec:inference} develops passive and active $x$-only inference,
Sec.~\ref{sec:matrix} gives the matrix generalization and the
configuration-dependent-mobility correction, and Sec.~\ref{sec:numerics}
provides the direct numerical test. Sections~\ref{sec:discussion} and
\ref{sec:conclusion} discuss the implications and conclude. Appendices collect
the entropy balance, reduced-current quadrature, projection proofs, and
numerical details.

\section{Model and stochastic dynamics}
\label{sec:model}

We consider one observed mechanical coordinate $x$ and $N$ hidden internal
phases $\bm\theta=(\theta_1,\dots,\theta_N)$. The observed coordinate is
periodic with spatial period $L$, or may be lifted to the real line with a
periodic interaction; each hidden phase lives on a circle,
$\theta_\alpha\in[0,2\pi)$. All coordinates are in contact with a single
isothermal reservoir at temperature $T$; we use energy units with Boltzmann's
constant set to unity, so that $T$ has the dimension of energy.

The coupling between the observed coordinate and the hidden phases is
reciprocal: it derives from a single shared interaction potential that depends
only on the combination
\begin{equation}
q=x-\sum_{\alpha=1}^{N}\ell_\alpha\theta_\alpha,
\label{eq:qdef}
\end{equation}
where $\ell_\alpha$ are coupling (gearing) lengths. Single-valuedness of the
interaction on the torus requires $\ell_\alpha=n_\alpha L/2\pi$ with integer
$n_\alpha$, so that $V(q)$ is $2\pi$-periodic in each hidden phase and a full
turn of a hidden phase advances $q$ by an integer number of spatial periods; the
derivations below use Eq.~(\ref{eq:qdef}) and the periodicity of the
interaction. The interaction energy is
\begin{equation}
U(x,\bm\theta)=V(q),\qquad V(q+L)=V(q),
\label{eq:Udef}
\end{equation}
and no further restriction is placed on $V$; it may be sinusoidal, asymmetric,
or any periodic function for which the overdamped diffusion is well defined.
Differentiating the shared potential gives the force on the observed coordinate
and the internal torque on each hidden phase,
\begin{equation}
F_x=-\partial_x U=-V'(q),\qquad
\tau_\alpha^{\mathrm{int}}=-\partial_{\theta_\alpha}U=\ell_\alpha V'(q),
\label{eq:forces}
\end{equation}
so that
\begin{equation}
\tau_\alpha^{\mathrm{int}}=-\ell_\alpha F_x.
\label{eq:reciprocity}
\end{equation}
Equation~(\ref{eq:reciprocity}) is a mechanical reciprocity identity, not a
linear-response relation: any force the internal coordinate exerts on the
observed coordinate is accompanied by the corresponding reaction torque on the
hidden coordinate. This structure distinguishes the present model from a one-way
hidden clock, in which the internal variable drives $x$ without back-reaction;
the consequences of the distinction are quantified in
Sec.~\ref{sec:discussion}.

The observed coordinate is opposed by a constant external load $f$, and each
hidden phase is driven by a constant nonconservative torque (affinity)
$\tau_\alpha$. The overdamped Langevin equations are
\begin{align}
\dot x&=\mu_x\bigl[-V'(q)-f\bigr]+\sqrt{2\mu_x T}\,\xi_x(t),
\label{eq:langevin_x}\\
\dot\theta_\alpha&=\mu_\alpha\bigl[\tau_\alpha+\ell_\alpha V'(q)\bigr]
+\sqrt{2\mu_\alpha T}\,\xi_\alpha(t),
\label{eq:langevin_theta}
\end{align}
with $\alpha=1,\dots,N$, where $\mu_x$ and $\mu_\alpha$ are mobilities and the
Gaussian white noises satisfy $\avg{\xi_i(t)}=0$ and
$\avg{\xi_i(t)\xi_j(t')}=\delta_{ij}\delta(t-t')$. Products defining heat and
work are taken in the Stratonovich convention, so that stochastic energetics
obeys the ordinary chain rule and the heat exchanged with the bath is the work
done by the systematic and frictional forces along a trajectory
\cite{Sekimoto2010}; this choice is essential because the interaction force
in Eqs.~(\ref{eq:langevin_x}) and (\ref{eq:langevin_theta}) is
state-dependent through $q$. The control parameters are the load $f$, the
affinities $\tau_\alpha$, the mobilities, the coupling lengths, and the
temperature; the observable is the mean current $J_x$ of the coordinate $x$.

\section{Fokker--Planck description, entropy balance, and power balance}
\label{sec:framework}

Let $p(x,\bm\theta,t)$ be the joint probability density on the torus. The
Fokker--Planck equation associated with
Eqs.~(\ref{eq:langevin_x})--(\ref{eq:langevin_theta}) is
\begin{equation}
\partial_t p=-\partial_x j_x-\sum_{\alpha=1}^{N}\partial_{\theta_\alpha}j_\alpha,
\label{eq:fp}
\end{equation}
with probability currents
\begin{align}
j_x&=\mu_x[-V'(q)-f]\,p-\mu_x T\,\partial_x p,
\label{eq:jx}\\
j_\alpha&=\mu_\alpha[\tau_\alpha+\ell_\alpha V'(q)]\,p
-\mu_\alpha T\,\partial_{\theta_\alpha}p.
\label{eq:jalpha}
\end{align}
The mean currents are the integrals
$J_x=\int j_x\,\dd x\,\dd\bm\theta$ and
$J_\alpha=\int j_\alpha\,\dd x\,\dd\bm\theta$, and we write the local mean
velocities as $\nu_x=j_x/p$ and $\nu_\alpha=j_\alpha/p$, so that
$J_x=\avg{\nu_x}$ and $J_\alpha=\avg{\nu_\alpha}$ under the stationary density.

The Gibbs--Shannon entropy $S=-\int p\ln p\,\dd x\,\dd\bm\theta$ evolves,
after substitution of Eq.~(\ref{eq:fp}) and integration by parts on the torus,
as
\begin{equation}
\frac{\dd S}{\dd t}=\ep-\dot h_{\mathrm{d}},
\label{eq:entropy_balance}
\end{equation}
where the entropy production rate is the manifestly non-negative
current-square functional
\begin{equation}
\ep=\int \dd x\,\dd\bm\theta
\left[\frac{j_x^2}{\mu_x T\,p}
+\sum_{\alpha=1}^{N}\frac{j_\alpha^2}{\mu_\alpha T\,p}\right]\ge0,
\label{eq:ep_full}
\end{equation}
and the entropy extraction (dissipation) rate is
\begin{equation}
\begin{split}
\dot h_{\mathrm{d}}=\frac{1}{T}\int \dd x\,\dd\bm\theta
\Big\{ &j_x[-V'(q)-f]\\
&+\sum_{\alpha=1}^{N}j_\alpha[\tau_\alpha+\ell_\alpha V'(q)]\Big\}.
\end{split}
\label{eq:hd}
\end{equation}
The derivation is given in Appendix~\ref{app:entropy}. At steady state
$\dd S/\dd t=0$, so $\ep^{\sstate}=\dot h_{\mathrm{d}}^{\sstate}$; this equality is a
balance, not a statement of equilibrium, and both rates can be positive.
Equation~(\ref{eq:ep_full}) already shows the essential point: $\ep=0$ requires
$j_x=0$ and every $j_\alpha=0$ pointwise, whereas the observable condition
$J_x=0$ constrains only the average of $\nu_x$.

A power balance follows from the time derivative of the mean interaction energy.
Using $\partial_x U=V'(q)$ and $\partial_{\theta_\alpha}U=-\ell_\alpha V'(q)$
and the stationarity of $\avg{U}$,
\begin{equation}
T\ep^{\sstate}=\sum_{\alpha=1}^{N}\tau_\alpha J_\alpha - fJ_x.
\label{eq:power}
\end{equation}
The affinities inject power $\sum_\alpha\tau_\alpha J_\alpha$, the load extracts
mechanical power $fJ_x$, and the difference is dissipated. Mechanical stall is
defined by the vanishing of the observable,
\begin{equation}
J_x(f_{\st})=0,
\label{eq:stall_def}
\end{equation}
at which the output power vanishes and Eq.~(\ref{eq:power}) reduces to
$T\ep^{\st}=\sum_\alpha\tau_\alpha J_\alpha^{\st}$. The input power through the
hidden cycles need not vanish, and it is the object of the remainder of the
analysis to determine how much of it the observed coordinate can constrain.

For completeness we record the projection statement in its general form. For an
overdamped diffusion $\dot z_i=b_i(z)+\sqrt{2D_i}\,\xi_i$ on a compact space,
the steady entropy production is
\begin{equation}
\ep=\int \dd z\sum_i \frac{(j_i^{\sstate})^2}{D_i\,p^{\sstate}},
\end{equation}
so that $\ep=0$ if and only if $j_i^{\sstate}(z)=0$ for all $i$ and all $z$. Any
observed current is a linear functional
$J_{\obs}=\int \dd z\,\bm a(z)\cdot\bm j^{\sstate}(z)$; the scalar condition
$J_{\obs}=0$ is implied by reversibility but does not imply it. Stall
is therefore weaker than reversibility for any coarse-grained observable. The
reciprocal model makes the resulting projected-current cancellation explicit and
quantitative.

\section{Reduction to the coupling coordinate and stall condition}
\label{sec:reduction}

Because the dynamics depends on $x$ and $\bm\theta$ only through the combination
$q$ in Eq.~(\ref{eq:qdef}), the difference
$\dot q=\dot x-\sum_\alpha\ell_\alpha\dot\theta_\alpha$ closes on $q$
(Appendix~\ref{app:reduction}):
\begin{equation}
\dot q=-M[V'(q)+\Lambda]+\sqrt{2MT}\,\eta_q(t),
\label{eq:qdot}
\end{equation}
with effective mobility and effective tilt
\begin{equation}
\begin{gathered}
M=\mu_x+\sum_{\alpha=1}^{N}\ell_\alpha^2\mu_\alpha,\\
\Lambda=\frac{\mu_x f+\sum_{\alpha=1}^{N}\ell_\alpha\mu_\alpha\tau_\alpha}{M},
\end{gathered}
\label{eq:M_Lambda}
\end{equation}
and $\eta_q$ a normalized white noise; the variance $2MT$ follows from the
independence of the constituent noises. The drift and diffusion depend on the
configuration only through $q$ and are invariant under the translations
$(x,\theta_\alpha)\mapsto(x+\sum_\beta\ell_\beta c_\beta,\ \theta_\alpha+c_\alpha)$
that leave $q$ fixed. Because the diffusion is elliptic on the compact torus it
is ergodic with a unique stationary density, and that density must share the
translation invariance of the generator; it is therefore constant along the
$N$-dimensional orbits and depends only on the transverse coordinate $q$,
$p^{\sstate}(x,\bm\theta)=(2\pi)^{-N}\rho(q)$ with $\int_0^L\rho\,\dd q=1$. The
reduced problem is a tilted periodic diffusion whose constant stationary current
$\mathcal J_q$ is given by the standard quadrature
\begin{widetext}
\begin{equation}
\mathcal J_q=\frac{MT\,[1-e^{\Lambda L/T}]}
{\int_0^L \dd q\,e^{-\psi(q)}\int_q^{q+L}\dd y\,e^{\psi(y)}},
\qquad \psi(q)=\frac{V(q)+\Lambda q}{T},
\label{eq:Jq}
\end{equation}
\end{widetext}
derived in Appendix~\ref{app:tilted}. The mean slip velocity is
$v_q=L\mathcal J_q$, and no assumption on the shape of $V$ has been made.

Averaging the Langevin equations gives the mean currents in terms of
$\avg{V'}=\int_0^L V'(q)\rho(q)\,\dd q$,
\begin{equation}
J_x=\mu_x[-\avg{V'}-f],\qquad
J_\alpha=\mu_\alpha[\tau_\alpha+\ell_\alpha\avg{V'}],
\label{eq:mean_currents}
\end{equation}
with the slip velocity $v_q=J_x-\sum_\alpha\ell_\alpha J_\alpha$. Eliminating
$\avg{V'}$ yields the mean-current identity
\begin{equation}
J_\alpha=\mu_\alpha(\tau_\alpha-\ell_\alpha f)
-\frac{\ell_\alpha\mu_\alpha}{\mu_x}J_x,
\label{eq:mean_identity}
\end{equation}
and, on substitution,
\begin{equation}
\begin{gathered}
J_x(f)=\frac{1}{K}\left[v_q(f)
+\sum_{\alpha=1}^{N}\ell_\alpha\mu_\alpha(\tau_\alpha-\ell_\alpha f)\right],\\
K=\frac{M}{\mu_x}=1+\frac{1}{\mu_x}\sum_{\alpha=1}^{N}\ell_\alpha^2\mu_\alpha.
\end{gathered}
\label{eq:Jx_from_vq}
\end{equation}
The dimensionless factor $K\ge1$ is the reciprocal amplification of the observed
mobility by the geared hidden coordinates. The stall load is the root
$J_x(f_{\st})=0$, i.e.
\begin{equation}
v_q(f_{\st})+\sum_{\alpha=1}^{N}\ell_\alpha\mu_\alpha(\tau_\alpha-\ell_\alpha f_{\st})=0,
\label{eq:stall_condition}
\end{equation}
which, with $v_q(f)$ fixed by Eq.~(\ref{eq:Jq}), defines $f_{\st}$ exactly for
any periodic $V$. At stall the hidden currents take the values
\begin{equation}
J_\alpha^{\st}=\mu_\alpha(\tau_\alpha-\ell_\alpha f_{\st}),
\label{eq:Jalpha_stall}
\end{equation}
which do not vanish in general.

\section{Local reciprocal-current identity and stall entropy production}
\label{sec:local}

The stationary density depends on $x$ and $\theta_\alpha$ only through $q$, so
$\partial_x\ln p^{\sstate}=\partial_q\ln\rho$ and
$\partial_{\theta_\alpha}\ln p^{\sstate}=-\ell_\alpha\partial_q\ln\rho$. Inserting
these into the current definitions gives the local velocities
\begin{align}
\nu_x(q)&=\mu_x[-V'(q)-f]-\mu_x T\,\partial_q\ln\rho(q),
\label{eq:nux}\\
\nu_\alpha(q)&=\mu_\alpha[\tau_\alpha+\ell_\alpha V'(q)]
+\ell_\alpha\mu_\alpha T\,\partial_q\ln\rho(q).
\label{eq:nua}
\end{align}
Eliminating $V'(q)$ and $\partial_q\ln\rho$ between
Eqs.~(\ref{eq:nux}) and (\ref{eq:nua}) yields the pointwise identity
\begin{equation}
\nu_\alpha(q)=\mu_\alpha(\tau_\alpha-\ell_\alpha f)
-\frac{\ell_\alpha\mu_\alpha}{\mu_x}\,\nu_x(q).
\label{eq:local_identity}
\end{equation}
Equation~(\ref{eq:local_identity}) holds for every $q$, every periodic $V$,
every drive, and every load; it is the local, current-level content of the
rank-one reciprocal coupling~(\ref{eq:reciprocity}), which enters through the
single combination $q$, and is stronger than the averaged
relation~(\ref{eq:mean_identity}). At stall $\avg{\nu_x}=J_x=0$, and
Eq.~(\ref{eq:local_identity}) becomes
\begin{equation}
\nu_\alpha^{\st}(q)=J_\alpha^{\st}-\frac{\ell_\alpha\mu_\alpha}{\mu_x}\,
\nu_x^{\st}(q).
\label{eq:local_stall}
\end{equation}

Define the positive current-square dissipation of the observed channel,
\begin{equation}
\Phi_x=\int_0^L \dd q\,\rho(q)\,\frac{\nu_x(q)^2}{\mu_x T}\ge0.
\label{eq:Phix}
\end{equation}
At stall $\avg{\nu_x^{\st}}=0$, but $\Phi_x^{\st}$ is positive whenever the local
observed current fails to vanish pointwise, i.e., whenever positive and negative
local currents cancel only after averaging. The quantity $\Phi_x$ is a
current-square dissipation carried by the $x$-channel, but it is a functional of
the full stationary state through the $q$-resolved local velocity $\nu_x(q)$ and
is not determined by the marginal statistics of $x$. The stationary
one-coordinate marginal is $p_x^{\sstate}(x)=1/L$, while its marginal
probability-current density
\begin{equation}
j_x^{\mathrm{marg}}(x)=\int \dd^N\theta\,j_x(x,\bm\theta)=\frac{J_x}{L}
\end{equation}
is spatially uniform and vanishes identically at stall. The stationary one-point
marginal and mean current of $x$ therefore reveal no dissipation at stall. A
static reconstruction of $\Phi_x^{\st}$ does require the $q$-resolved local
velocity $\nu_x(q)$, obtained by resolving the coupling coordinate or by
calibrated force measurements. This static limitation is not a dynamical
no-go: Sec.~\ref{sec:inference} proves that a controlled fluctuation--response
measurement of $x(t)$ alone reconstructs $\Phi_x$ exactly for the present
constant-mobility reciprocal model, while passive time-reversal statistics give
a generally weaker lower bound. With this distinction understood, the full
entropy production
(\ref{eq:ep_full}) expressed through the local velocities is
\begin{equation}
\ep=\int_0^L \dd q\,\rho\left[\frac{\nu_x^2}{\mu_x T}
+\sum_\alpha \frac{\nu_\alpha^2}{\mu_\alpha T}\right].
\end{equation}
Squaring Eq.~(\ref{eq:local_stall}) and averaging with $\rho$ removes the cross
term because $\avg{\nu_x^{\st}}=0$,
\begin{equation}
\avg{(\nu_\alpha^{\st})^2}=(J_\alpha^{\st})^2
+\frac{\ell_\alpha^2\mu_\alpha^2}{\mu_x^2}\avg{(\nu_x^{\st})^2},
\end{equation}
and substitution gives the exact stall entropy production
\begin{equation}
\boxed{\;\ep^{\st}=K\,\Phi_x^{\st}
+\sum_{\alpha=1}^{N}\frac{(J_\alpha^{\st})^2}{\mu_\alpha T}\;}
\label{eq:stall_ep_1}
\end{equation}
with $K$ as in Eq.~(\ref{eq:Jx_from_vq}). The result is exact for any periodic
$V$. The first term is the reciprocal local-current contribution, positive when
observed local currents survive stall; the second is the hidden mean-cycle
contribution, positive when any hidden current remains.
Equation~(\ref{eq:stall_ep_1}) vanishes only if $\Phi_x^{\st}=0$, which requires
$\nu_x^{\st}(q)=0$ for all $q$, and every $J_\alpha^{\st}=0$; by
Eq.~(\ref{eq:local_stall}) these together force $\nu_\alpha^{\st}(q)=0$ for all
$q$, i.e., the full current field vanishes. Stall coincides with equilibrium only
in that full-current limit.

Dimensional consistency is direct: $\mu_x T$ has dimensions of
$(\mathrm{length})^2/\mathrm{time}$ and $\nu_x^2$ has dimensions of
$(\mathrm{length}/\mathrm{time})^2$, so $\nu_x^2/(\mu_x T)$ has dimensions of
$\mathrm{time}^{-1}$; the hidden term $J_\alpha^2/(\mu_\alpha T)$ is likewise a
rate, so both terms of Eq.~(\ref{eq:stall_ep_1}) are entropy production rates.

\section{Reciprocal and orthogonal decomposition, closure, and bound}
\label{sec:decomp}

The exact result~(\ref{eq:stall_ep_1}) contains all hidden currents, but the
observed coordinate cannot resolve them separately. To isolate the part it does
fix, introduce the collective reciprocal current
\begin{equation}
J_\parallel=\sum_{\alpha=1}^{N}\ell_\alpha J_\alpha,
\label{eq:Jpar_def}
\end{equation}
the projection of the hidden-current vector onto the coupling direction.
Combining the stall power balance $T\ep^{\st}=\sum_\alpha\tau_\alpha
J_\alpha^{\st}$ with $\tau_\alpha=\ell_\alpha f_{\st}+J_\alpha^{\st}/\mu_\alpha$
from Eq.~(\ref{eq:Jalpha_stall}) gives
$T\ep^{\st}=f_{\st}J_\parallel^{\st}+\sum_\alpha (J_\alpha^{\st})^2/\mu_\alpha$,
while multiplying Eq.~(\ref{eq:stall_ep_1}) by $T$ gives
$T\ep^{\st}=TK\Phi_x^{\st}+\sum_\alpha (J_\alpha^{\st})^2/\mu_\alpha$. The hidden
quadratic term cancels between the two expressions, leaving the closure relation
\begin{equation}
\boxed{\;f_{\st}\,J_\parallel^{\st}=T K\,\Phi_x^{\st}\;}
\label{eq:closure}
\end{equation}
so that, for $f_{\st}\neq0$, the collective hidden current is fixed by the
observed-channel dissipation and the stall load,
\begin{equation}
J_\parallel^{\st}=\frac{TK\,\Phi_x^{\st}}{f_{\st}}.
\label{eq:Jpar_recon}
\end{equation}
Only this one scalar combination is fixed by the observed channel, and, as noted
above, evaluating the right-hand side requires $\Phi_x^{\st}$, hence resolution
of the coupling coordinate rather than displacement statistics alone.

To see what remains hidden, minimize the hidden dissipation
$\sum_\alpha J_\alpha^2/\mu_\alpha$ at fixed $J_\parallel$
(Appendix~\ref{app:decomp}). With
\begin{equation}
B=\sum_{\alpha=1}^{N}\ell_\alpha^2\mu_\alpha=\mu_x(K-1),
\end{equation}
the minimizer is $J_\alpha^{\parallel}=(\mu_\alpha\ell_\alpha/B)J_\parallel$, and
any hidden-current vector decomposes as
$J_\alpha=J_\alpha^{\parallel}+J_\alpha^{\perp}$ with
$\sum_\alpha\ell_\alpha J_\alpha^{\perp}=0$. In the mobility-weighted metric the
decomposition is orthogonal,
\begin{equation}
\sum_{\alpha}\frac{J_\alpha^2}{\mu_\alpha}
=\frac{J_\parallel^2}{B}+\sum_\alpha\frac{(J_\alpha^{\perp})^2}{\mu_\alpha},
\label{eq:pythag}
\end{equation}
so that the stall entropy production splits into a reconstructable reciprocal
part and an orthogonal part,
\begin{equation}
\begin{gathered}
\ep^{\st}=K\Phi_x^{\st}+\frac{(J_\parallel^{\st})^2}{BT}
+\mathcal D_\perp^{\st},\\
\mathcal D_\perp^{\st}=\sum_{\alpha=1}^{N}
\frac{(J_\alpha^{\perp,\st})^2}{\mu_\alpha T}\ge0.
\end{gathered}
\label{eq:stall_perp}
\end{equation}
Using Eq.~(\ref{eq:Jpar_recon}) and $B=\mu_x(K-1)$, for $f_{\st}\neq0$,
\begin{equation}
\boxed{\;
\begin{aligned}
\ep^{\st}={}&K\Phi_x^{\st}
+\frac{T K^2 (\Phi_x^{\st})^2}{\mu_x(K-1)f_{\st}^2}\\
&+\mathcal D_\perp^{\st}
\end{aligned}
\;}
\label{eq:stall_final}
\end{equation}
and, since $\mathcal D_\perp^{\st}\ge0$,
\begin{equation}
\ep^{\st}\ge K\Phi_x^{\st}
+\frac{T K^2 (\Phi_x^{\st})^2}{\mu_x(K-1)f_{\st}^2},
\label{eq:bound_K}
\end{equation}
with equality if and only if no hidden current circulates orthogonally to the
coupling direction. Equations~(\ref{eq:stall_final}) and (\ref{eq:bound_K})
are the central results. The reciprocal part is fixed by the observed
current-square dissipation $\Phi_x^{\st}$, the stall load $f_{\st}$, and the
factor $K$; the orthogonal part $\mathcal D_\perp^{\st}$ is a genuine
thermodynamic cost that is invisible to the observed coordinate. Two motors with
identical observed dynamics and identical reciprocal projection but different
$\mathcal D_\perp^{\st}$ produce different total entropy at stall. This is the
no-go content: a single observed coordinate can determine at most one scalar
combination of hidden currents, and for $N\ge2$ there exist hidden loops that
leave the observed coordinate unchanged.

Equations~(\ref{eq:Jpar_recon})--(\ref{eq:bound_K}) require $f_{\st}\neq0$,
since they divide by $f_{\st}$. The limit $f_{\st}=0$ is not a loss of
information but a collapse of the reciprocal channel. The closure~(\ref{eq:closure})
then forces $\Phi_x^{\st}=0$, hence $\nu_x^{\st}(q)=0$ for all $q$; by
Eq.~(\ref{eq:nux}) this makes $\rho\propto e^{-V/T}$ the equilibrium weight of
the interaction, which is stationary for the reduced diffusion~(\ref{eq:qdot})
only at zero tilt, $\Lambda=0$. For nontrivial $V$ this gives $v_q=0$ and
$J_\parallel^{\st}=0$. At zero stall load the reciprocal current is therefore
silent, and only the orthogonal hidden currents $\mathcal D_\perp^{\st}$ can
remain.

When the factor $K$ is not independently
calibrated, the bound~(\ref{eq:bound_K}) can be minimized over $K>1$. Writing
$a=T\Phi_x^{\st}/(\mu_x f_{\st}^2)$ and $K=1+s$, the bracket becomes
$1+2a+(1+a)s+a/s$, minimized at $s_*=\sqrt{a/(1+a)}$, giving the model-class
lower bound
\begin{equation}
\ep^{\st}\ge \Phi_x^{\st}\bigl(\sqrt{1+a}+\sqrt a\bigr)^2,
\label{eq:unknownK}
\end{equation}
which is strictly stronger than the visible-channel estimate
$\ep^{\st}\ge\Phi_x^{\st}$ whenever $\Phi_x^{\st}>0$.

A single hidden coordinate is the special case $N=1$. With hidden phase
$\theta$, coupling $\ell$, mobility $\mu_\theta$, and $B=\ell^2\mu_\theta$, the
hidden space has no orthogonal direction, so $\mathcal D_\perp^{\st}=0$ and the
bound~(\ref{eq:bound_K}) is an equality,
\begin{equation}
\begin{gathered}
\ep^{\st}=K\Phi_x^{\st}
+\frac{TK^2(\Phi_x^{\st})^2}{\mu_x(K-1)f_{\st}^2},\\
J_\theta^{\st}=\frac{TK\Phi_x^{\st}}{\ell f_{\st}}.
\end{gathered}
\label{eq:single_cycle}
\end{equation}
The single-cycle motor thus has no irreducible orthogonal cost; its hidden
current is fixed entirely by the observed-channel dissipation and stall load
through Eq.~(\ref{eq:single_cycle}), subject to the same requirement that
$\Phi_x^{\st}$ be resolved. The irreducible term $\mathcal D_\perp^{\st}$ appears
only for $N\ge2$.

\section{An exactly solvable nonreciprocal realization}
\label{sec:solvable}

The general results above rest on rank-one reciprocal coupling. To display the
two elementary sources of stall dissipation in closed form, independently of that
structure, we solve a minimal nonreciprocal model in which the observed
coordinate is driven by an arbitrary periodic active force and the hidden phase
advances as an autonomous clock. Because the model is nonreciprocal it does not
test the closure~(\ref{eq:closure}); rather, it isolates local force
fluctuations and hidden circulation as the two contributions to $\ep^{\st}$, and
serves as an analytical check on the positivity structure of
Sec.~\ref{sec:local}. The dynamics is
\begin{align}
\dot x&=\mu[F(\theta)-f]+\sqrt{2\mu T}\,\xi_x(t),\label{eq:solv_x}\\
\dot\theta&=\Omega+\sqrt{2D_\theta}\,\xi_\theta(t),\label{eq:solv_theta}
\end{align}
with $x$ on a ring of period $L$, $\theta$ on a ring of period $2\pi$, $\mu$ the
observed mobility, $F(\theta+2\pi)=F(\theta)$ an arbitrary periodic active
force, $\Omega$ the hidden drift, and $D_\theta$ the hidden diffusion
coefficient; if desired one may write $D_\theta=\mu_\theta T_\theta$ and
$\Omega=\mu_\theta\tau_\theta$. Define the phase average
$\bar F=(2\pi)^{-1}\int_0^{2\pi}F\,\dd\theta$ and the phase variance
$V_F=(2\pi)^{-1}\int_0^{2\pi}(F-\bar F)^2\,\dd\theta=\Var[F]$.

Because the drifts are independent of $x$, the generator is translation
invariant in $x$, and the stationary density is uniform (Appendix~\ref{app:uniform}),
\begin{equation}
p^{\sstate}(x,\theta)=\frac{1}{2\pi L},
\end{equation}
with steady currents
\begin{equation}
j_x^{\sstate}(\theta)=\frac{\mu[F(\theta)-f]}{2\pi L},\qquad
j_\theta^{\sstate}=\frac{\Omega}{2\pi L}.
\label{eq:solv_currents}
\end{equation}
The observed current is the phase average
\begin{equation}
J_x(f)=\mu(\bar F-f),
\label{eq:solv_Jx}
\end{equation}
so the stall load is the phase average of the active force,
\begin{equation}
f_{\st}=\bar F,
\label{eq:solv_stall}
\end{equation}
at which the output power $P_{\mathrm{out}}=fJ_x$ vanishes. The local current at
stall is $j_x^{\sstate}(\theta;f_{\st})=\mu[F(\theta)-\bar F]/(2\pi L)$; its phase
average vanishes but it is nonzero pointwise unless $F$ is constant.

Evaluating Eq.~(\ref{eq:ep_full}) on the uniform steady state gives the entropy
production for arbitrary load,
\begin{equation}
\ep^{\sstate}(f)=\frac{\mu}{T}(\bar F-f)^2+\frac{\mu}{T}V_F
+\frac{\Omega^2}{D_\theta},
\label{eq:solv_ep_load}
\end{equation}
where the decomposition $F-f=(F-\bar F)+(\bar F-f)$ separates the mean-current
contribution from the variance contribution (Appendix~\ref{app:uniform}). At
stall the first term vanishes and
\begin{equation}
\boxed{\;\ep^{\st}=\frac{\mu}{T}\,\Var[F]+\frac{\Omega^2}{D_\theta}\;}
\label{eq:solv_ep_stall}
\end{equation}
The observed current~(\ref{eq:solv_Jx}) depends only on $\bar F$, whereas the
stall dissipation~(\ref{eq:solv_ep_stall}) depends on the force variance and the
hidden drift, neither of which enters $J_x(f)$. Thus different force profiles
with the same mean, and any hidden drift, share the same current--load curve and
the same stall load but produce different stall dissipation; the mean current
alone cannot determine $\ep^{\st}$. This is the solvable-model form of the no-go
statement of Sec.~\ref{sec:decomp}, and the entropy extraction rate can be
verified independently to equal $\ep^{\sstate}$ at steady state
(Appendix~\ref{app:uniform}). Reversible stall requires $V_F=0$ and $\Omega=0$
simultaneously, i.e., a phase-independent force and a non-circulating clock, in
which case the load $f=\bar F$ balances a constant force and all currents
vanish.

\begin{figure}[t]
\centering
\includegraphics[width=0.98\linewidth]{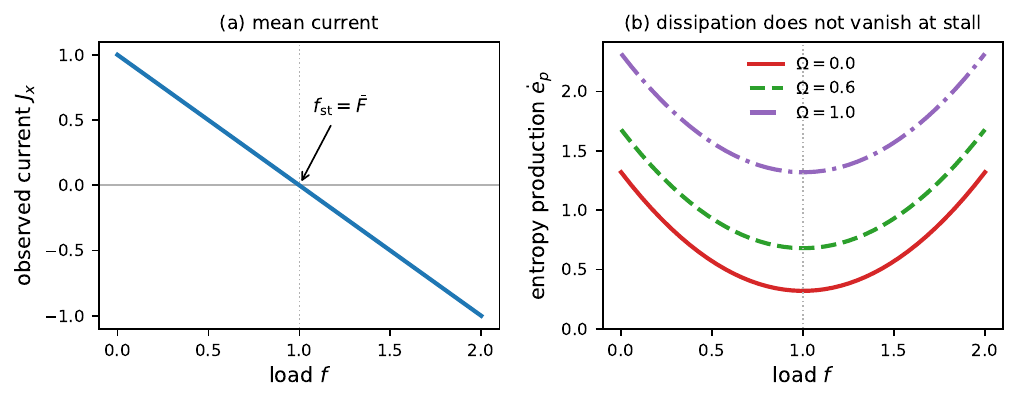}
\caption{Exact analytic behavior of the solvable hidden-phase model,
Eqs.~(\ref{eq:solv_Jx}) and (\ref{eq:solv_ep_load}), for
$F(\theta)=F_0+A\cos\theta$ with $F_0=1$, $A=0.8$, and $\mu=T=D_\theta=1$
(energy units). (a) The observed mean current $J_x(f)=\mu(\bar F-f)$ vanishes at
the stall load $f_{\st}=\bar F=F_0$ (dotted line). (b) The entropy production
rate $\ep(f)$ for three hidden drifts $\Omega$; at the stall load the
dissipation does not vanish but equals $\mu V_F/T+\Omega^2/D_\theta$
($=0.32,0.68,1.32$ for $\Omega=0,0.6,1.0$). Curves are evaluations of the
closed-form expressions, not simulations.}
\label{fig:stall}
\end{figure}

Figure~\ref{fig:stall} evaluates Eqs.~(\ref{eq:solv_Jx}) and
(\ref{eq:solv_ep_load}) for the representative choice
$F(\theta)=F_0+A\cos\theta$, for which $\bar F=F_0$ and $V_F=A^2/2$. The observed
current crosses zero at $f_{\st}=F_0$, where the entropy production retains the
finite value $\mu V_F/T+\Omega^2/D_\theta$. The curves are direct evaluations of
the closed-form expressions, not stochastic simulations, and are shown to make
the analytic statement concrete.

\section{Heat-engine bookkeeping with hidden heat exchange and the Carnot bound}
\label{sec:heat}

The preceding results are isothermal statements about entropy production. When
the motor is embedded between a hot reservoir at temperature $T_h$ and a cold
reservoir at $T_c<T_h$, the same distinction between mechanical stall and
reversibility reappears at the level of reservoir heat currents. We adopt the
heat-engine convention
in which $q_h>0$ is heat taken from the hot reservoir and $q_c>0$ is heat given
to the cold reservoir, both positive magnitudes in the forward regime
\cite{Asfaw2005EnergeticsSimpleMicroscopic,Taye2017IrreversibleBrownianHeat}.

For the visible mechanical step of length $L_0$ against load $f$, with barrier
energy scale $U_0$, the standard Brownian-engine accounting assigns
\begin{equation}
q_h^{(0)}=U_0+\frac{fL_0}{2},\qquad q_c^{(0)}=U_0-\frac{fL_0}{2},
\label{eq:step_heat}
\end{equation}
so that the visible work per step is $q_h^{(0)}-q_c^{(0)}=fL_0$. With net
forward step current $J$, the net visible heat rates are
$\dot q_h^{\vis}=Jq_h^{(0)}$ and $\dot q_c^{\vis}=Jq_c^{(0)}$, and the output
power is $P_{\mathrm{out}}=JfL_0$; these are net rates, since forward and
backward transitions each exchange heat and only their signed difference,
proportional to $J$, appears here. Hidden degrees of freedom enter through
reservoir-resolved heat rates $\dot q_h^{\hid}$ and $\dot q_c^{\hid}$, giving the
total rates
\begin{equation}
\begin{gathered}
\dot q_h=J\!\left(U_0+\frac{fL_0}{2}\right)+\dot q_h^{\hid},\\
\dot q_c=J\!\left(U_0-\frac{fL_0}{2}\right)+\dot q_c^{\hid}.
\end{gathered}
\label{eq:total_heat}
\end{equation}
Energy conservation at steady state reads
$\dot q_h-\dot q_c=P_{\mathrm{out}}+\dot W_{\hid}$, where $\dot W_{\hid}$ is the
net work exchanged through the hidden degrees of freedom. Throughout this section
we take the hidden channel to carry heat without performing net work,
$\dot W_{\hid}=0$, so that
\begin{equation}
\dot q_h^{\hid}=\dot q_c^{\hid}\equiv\dot q_{\mathrm{leak}},\qquad
P_{\mathrm{out}}=\dot q_h-\dot q_c;
\label{eq:nowork}
\end{equation}
a hidden channel that also performs work, such as a chemical drive, would add an
explicit affinity term to Eq.~(\ref{eq:ep_engine}) and lies outside this
bookkeeping.

The total entropy production of the two-reservoir engine is
\begin{equation}
\ep=\frac{\dot q_c}{T_c}-\frac{\dot q_h}{T_h}\ge0,
\label{eq:ep_engine}
\end{equation}
which separates into a visible and a hidden contribution,
$\ep=\ep^{\vis}+\ep^{\hid}$, with
$\ep^{\hid}=\dot q_{\mathrm{leak}}(1/T_c-1/T_h)$. The efficiency is
$\eta=P_{\mathrm{out}}/\dot q_h=1-\dot q_c/\dot q_h$; eliminating $\dot q_c$ with
Eq.~(\ref{eq:ep_engine}) gives the single relation
\begin{equation}
\eta=\eta_C-\frac{T_c\,\ep}{\dot q_h},\qquad \eta_C=1-\frac{T_c}{T_h}.
\label{eq:eta_carnot}
\end{equation}
The relation presumes $\dot q_h>0$; when no heat is drawn the efficiency is
undefined. At every finite operating point with $\dot q_h>0$, positive entropy
production implies $\eta<\eta_C$. In a regular overdamped or Markov model
satisfying local detailed balance, zero entropy production implies vanishing
irreversible probability currents, so Carnot efficiency can be approached only in
the reversible zero-current and zero-power limit. Equation~(\ref{eq:ep_full})
establishes the zero-current implication for the isothermal model of
Secs.~\ref{sec:model}--\ref{sec:solvable}.

Mechanical stall is $J=0$. Then $P_{\mathrm{out}}^{\st}=0$ and the net visible
heat rates vanish, while Eq.~(\ref{eq:total_heat}) leaves the hidden rate finite,
$\dot q_h^{\st}=\dot q_c^{\st}=\dot q_{\mathrm{leak}}^{\st}$. If heat is still
drawn from the hot reservoir the stall efficiency is $\eta_{\st}=0$ and the
entropy production is positive,
\begin{equation}
\begin{gathered}
\ep^{\st}=\dot q_{\mathrm{leak}}^{\st}\!\left(\frac{1}{T_c}-\frac{1}{T_h}\right),\\
\dot q_{\mathrm{leak}}^{\st}=\frac{T_c}{\eta_C}\,\ep^{\st},
\end{gathered}
\label{eq:leak}
\end{equation}
so a stalled engine is reversible only if the leak vanishes. This reproduces,
at the level of reservoir bookkeeping, the isothermal conclusion of
Secs.~\ref{sec:local}--\ref{sec:decomp}: mechanical silence does not imply
reversibility. The reduction of the endoreversible or Curzon--Ahlborn efficiency
by hidden heat transfer, $\eta=\eta_*/(1+\dot q_h^{\hid}/\dot q_h^{\vis})$ for
any visible reference efficiency $\eta_*$, follows immediately and connects to
finite-time analyses of such engines
\cite{CurzonAhlborn1975,Taye2025CurzonAhlbornType,Taye2025ThermodynamicFeaturesHeat}.
Unlike the reciprocal case, however, an arbitrary hidden heat rate is not
reconstructable from the observed current: a leak orthogonal to the observed
motion can be added without changing $J$, so calorimetry or a known coupling
structure is required to fix it.

\section{Inference from the observed trajectory and response}
\label{sec:inference}

The previous sections used the $q$-resolved current to evaluate $\Phi_x$. Here
we distinguish two experimentally different questions. First, can a controlled
perturbation of the observed coordinate reconstruct the reciprocal
dissipation? Second, what can be inferred passively from an unperturbed
trajectory? For the present reciprocal model the first question has an exact
answer, whereas the second gives a lower bound.

\subsection{Exact fluctuation--response reconstruction}

Let $\dot Q_x$ denote the stationary heat-dissipation rate into the bath through
the observed coordinate, with the sign convention of
Eq.~(\ref{eq:hd}),
\begin{equation}
\begin{aligned}
\dot Q_x={}&\int \dd x\,\dd\bm\theta\,
 j_x(x,\bm\theta)[-V'(q)-f]\\
={}&\int_0^L\dd q\,\rho(q)\nu_x(q)[-V'(q)-f].
\end{aligned}
\label{eq:Qx_def}
\end{equation}
Equation~(\ref{eq:nux}) gives
\begin{equation}
\frac{\dot Q_x}{T}=\Phi_x+\mathcal I_x,
\qquad
\mathcal I_x=\int_0^L\nu_x(q)\rho'(q)\,\dd q.
\label{eq:Qx_phi_pre}
\end{equation}
The correction $\mathcal I_x$ vanishes because of the rank-one kinematics. Define
\begin{equation}
A(f)=\sum_{\alpha=1}^{N}\ell_\alpha\mu_\alpha
(\tau_\alpha-\ell_\alpha f).
\label{eq:Adef}
\end{equation}
Using Eq.~(\ref{eq:local_identity}), the local velocity of the coupling
coordinate is
\begin{equation}
\nu_q(q)=\nu_x(q)-\sum_\alpha\ell_\alpha\nu_\alpha(q)
=K\nu_x(q)-A(f).
\label{eq:nuq_nux}
\end{equation}
Stationarity of the reduced diffusion implies
$\rho(q)\nu_q(q)=\mathcal J_q$, so
\begin{equation}
\nu_x(q)=\frac{1}{K}\left[\frac{\mathcal J_q}{\rho(q)}+A(f)\right].
\label{eq:nux_Jq}
\end{equation}
Substitution into Eq.~(\ref{eq:Qx_phi_pre}) gives
\begin{equation}
\mathcal I_x=\frac{\mathcal J_q}{K}
\int_0^L\partial_q\ln\rho\,\dd q
+\frac{A}{K}\int_0^L\partial_q\rho\,\dd q=0,
\end{equation}
because $\rho$ is positive and periodic. Hence
\begin{equation}
\boxed{\;\frac{\dot Q_x}{T}=\Phi_x\;}
\label{eq:Qx_equals_phi}
\end{equation}
for every load, not only at stall. The equality is stronger than the generic
coordinate-wise entropy balance; it follows from the single reduced current and
periodicity of the reciprocal model.

To express Eq.~(\ref{eq:Qx_equals_phi}) using $x(t)$ alone, add a weak probe
force $h(t)$ to Eq.~(\ref{eq:langevin_x}), so that the observed drift contains
$+\mu_xh(t)$. Define the connected velocity correlation and causal response
\begin{align}
C_{vv}(t)&=\avg{[\dot x(t)-J_x][\dot x(0)-J_x]},\\
R_v(t-s)&=\left.\frac{\delta\avg{\dot x(t)}}{\delta h(s)}\right|_{h=0},
\qquad t>s,
\end{align}
and let tildes denote Fourier transforms. The Harada--Sasa equality
\cite{Harada2005} gives
\begin{equation}
\dot Q_x=\frac{J_x^2}{\mu_x}
+\frac{1}{\mu_x}\int_{-\infty}^{\infty}\frac{\dd\omega}{2\pi}
\left[\widetilde C_{vv}(\omega)
-2T\,\mathrm{Re}\,\widetilde R_v(\omega)\right].
\label{eq:HS}
\end{equation}
Combining Eqs.~(\ref{eq:Qx_equals_phi}) and (\ref{eq:HS}) yields the exact
$x$-only reconstruction
\begin{equation}
\begin{aligned}
\Phi_x={}&\frac{J_x^2}{\mu_xT}\\
&+\frac{1}{\mu_xT}\int_{-\infty}^{\infty}\frac{\dd\omega}{2\pi}
\left[\widetilde C_{vv}(\omega)
-2T\,\mathrm{Re}\,\widetilde R_v(\omega)\right].
\end{aligned}
\label{eq:Phi_HS}
\end{equation}
At stall the mean-current term vanishes. Experimentally, the singular
high-frequency pieces of the correlation and response must be evaluated with
the same temporal resolution; their difference has a finite continuum limit.
Substitution of Eq.~(\ref{eq:Phi_HS}) into
Eqs.~(\ref{eq:Jpar_recon}) and (\ref{eq:bound_K}) converts the reciprocal
current and the total-dissipation bound into quantities obtained from the stall
load, calibrated mobilities, and the measured correlation--response violation
of $x(t)$ alone.

\subsection{Passive time-reversal bound}

A passive trajectory does not provide the response function, but its arrow of
time still supplies a model-independent bound. Let
$\mathcal P_x^{(\mathcal T)}[x_{[0,\mathcal T]}]$ be the path measure of the
observed coordinate over a duration $\mathcal T$, and let
$\mathcal P_x^{(\mathcal T),\dagger}$ be the measure of the time-reversed path.
The observed irreversibility rate is
\begin{equation}
\dot e_{\mathrm{p},x}^{\mathrm{TR}}=
\lim_{\mathcal T\to\infty}\frac{1}{\mathcal T}
D_{\mathrm{KL}}\!\left(
\mathcal P_x^{(\mathcal T)}\middle\|\mathcal P_x^{(\mathcal T),\dagger}
\right).
\label{eq:path_KL}
\end{equation}
For the full Markov process the corresponding relative-entropy rate equals the
steady entropy production under local detailed balance. Projection onto $x$ is
a stochastic map, so the data-processing inequality gives
\begin{equation}
0\le \dot e_{\mathrm{p},x}^{\mathrm{TR}}\le\ep .
\label{eq:path_bound}
\end{equation}
This bound can remain positive when $J_x=0$ and therefore detects irreversible
fluctuations with no mean flow \cite{Gaspard2004,RoldanParrondo2010}. It is not,
however, a complete reconstruction: an orthogonal hidden circulation can leave
the entire $x$ path measure unchanged, in which case
$\dot e_{\mathrm{p},x}^{\mathrm{TR}}=0$ while $\mathcal D_\perp^{\st}>0$. Equations
(\ref{eq:Phi_HS}) and (\ref{eq:path_bound}) thus distinguish active-response
inference, which is exact for the reciprocal channel, from passive inference,
which is universally valid but generally incomplete.

\section{Operational reconstruction from finite data}
\label{sec:protocol}

Equations~(\ref{eq:Phi_HS}) and (\ref{eq:Jpar_recon}) are continuum
identities, whereas an experiment or simulation records a finite, filtered time
series. This section turns the identities into a practical protocol and states
which conclusions remain exact after finite-bandwidth estimation.

\subsection{Calibration, stall location, and response measurement}

The observed mobility $\mu_x$ and bath temperature $T$ should first be
calibrated independently, or from a reference state in which the applied probe
is known. The load is then varied while all hidden drives are held fixed, and
$J_x(f)$ is measured on both sides of its zero. A local interpolation gives
$f_{\st}$ and its uncertainty without assuming that the current--load curve is
linear away from stall. The correlation and response measurements must be made
at this same load and with the same sampling and filtering.

For the unperturbed record, one estimates the connected velocity spectrum
$\widetilde C_{vv}(\omega)$. A weak zero-mean probe $h(t)$, composed for example
of separated sinusoids or a broadband signal, gives the complex linear response
$\widetilde R_v(\omega)$. Linearity should be tested by repeating the
measurement at a smaller probe amplitude. Because ideal overdamped velocities
contain white-noise singularities, the correlation and response must be
processed with the same finite-difference rule, detector transfer function, and
frequency window. The Harada--Sasa difference is then filter-consistent even
when the two terms are not separately well behaved in the continuum limit.

For a symmetric angular-frequency cutoff $\Omega$, define the measured
functional
\begin{equation}
\begin{aligned}
\Phi_x^{(\Omega)}={}&\frac{J_x^2}{\mu_xT}\\
&+\frac{1}{\mu_xT}\int_{-\Omega}^{\Omega}
\frac{\dd\omega}{2\pi}
\left[\widetilde C_{vv}(\omega)
-2T\,\mathrm{Re}\,\widetilde R_v(\omega)\right].
\end{aligned}
\label{eq:Phi_cutoff}
\end{equation}
Under the regularity conditions of the overdamped model,
$\Phi_x^{(\Omega)}\to\Phi_x$ as the resolved bandwidth grows. The omitted-tail
identity is exact,
\begin{equation}
\Phi_x-\Phi_x^{(\Omega)}=
\frac{1}{\mu_xT}\int_{|\omega|>\Omega}\frac{\dd\omega}{2\pi}
\left[\widetilde C_{vv}-2T\,\mathrm{Re}\,\widetilde R_v\right],
\label{eq:tail_identity}
\end{equation}
but the integrand need not have a fixed sign. A finite-bandwidth value is
therefore an estimator, not automatically a rigorous lower bound. Convergence
should be demonstrated by increasing $\Omega$ until the result is stable within
its statistical error, or by bounding the unresolved absolute tail.

\subsection{Inference and one-sided uncertainty}

When the coupling factor $K$ is calibrated and $f_{\st}\ne0$, the plug-in
estimates are
\begin{align}
\widehat J_{\parallel}^{(\Omega)}&=
\frac{TK\Phi_x^{(\Omega)}}{f_{\st}},
\label{eq:Jpar_cutoff}\\
\widehat e_{\mathrm{p,rec}}^{(\Omega)}&=
K\Phi_x^{(\Omega)}+
\frac{TK^2[\Phi_x^{(\Omega)}]^2}
{\mu_x(K-1)f_{\st}^2}.
\label{eq:ep_cutoff}
\end{align}
Equation~(\ref{eq:ep_cutoff}) estimates the reciprocal lower bound, not the
orthogonal cost. To preserve a one-sided thermodynamic statement in the
presence of sampling error, one may replace $\Phi_x$ by a nonnegative lower
confidence endpoint $\Phi_x^{\mathrm L}$. Since the right-hand side of
Eq.~(\ref{eq:bound_K}) is monotone in $\Phi_x\ge0$, the confidence-controlled
statement is
\begin{equation}
\ep^{\st}\ge
K\Phi_x^{\mathrm L}+
\frac{TK^2(\Phi_x^{\mathrm L})^2}
{\mu_x(K-1)f_{\st}^2}.
\label{eq:confidence_bound}
\end{equation}
If $K$ is unknown, the model-class bound~(\ref{eq:unknownK}) can be
used instead; it requires only $T$, $\mu_x$, $f_{\st}$, and the reconstructed
$\Phi_x$. The reconstruction has three immediate consistency checks:
$\Phi_x\ge0$; $f_{\st}J_\parallel^{\st}=TK\Phi_x^{\st}\ge0$; and the measured
correlation--response violation must disappear in the reversible limit in which
all stationary currents vanish.

\subsection{Hierarchy of observability}

Table~\ref{tab:observability} summarizes the information gained by increasingly
rich observations. Static displacement data determine neither $\Phi_x$ nor the
hidden cost at stall. Passive path data give a universal irreversibility bound.
An active response measurement reconstructs the entire reciprocal channel in
the constant-mobility rank-one model. Recording additional internal markers
adds rows to the effective coupling matrix and reduces the invisible null
space. Only a state description spanning all current-carrying directions fixes
the total entropy production without a residual no-go sector.

\begin{table*}[t]
\caption{Observation hierarchy at mechanical stall. ``Exact'' refers to the
stated constant-mobility reciprocal model; no method based on a single observed
coordinate can recover a current wholly in the coupling null space.}
\label{tab:observability}
\begin{ruledtabular}
\begin{tabular}{p{0.18\textwidth}p{0.23\textwidth}p{0.27\textwidth}p{0.22\textwidth}}
Observation & Measured data & Thermodynamic information & Remaining ambiguity\\
\hline
Static $x$ & $p_x^{\sstate}(x)$ and $J_x$ & No dissipation signal at stall & Local cancellation and all hidden currents\\
Passive $x(t)$ & Forward/reversed path statistics & $\dot e_{\mathrm{p},x}^{\mathrm{TR}}\le\ep$ & Dynamically invisible hidden loops\\
Active $x(t)$ & $C_{vv}$ and $R_v$ & Exact $\Phi_x$, $J_\parallel^{\st}$, and reciprocal lower bound & Orthogonal hidden dissipation\\
$x(t)$ plus internal markers & Augmented trajectory and response data & More components of the projected hidden current & Nullity of the augmented coupling matrix\\
Full state & All probability currents and reservoir labels & Exact total entropy production & None within the model\\
\end{tabular}
\end{ruledtabular}
\end{table*}

\section{Matrix coupling and configuration-dependent mobilities}
\label{sec:matrix}

\subsection{Several observed interaction coordinates}

Consider $m$ observed coordinates $\bm x\in\mathbb T^m$ and $N$ hidden phases
$\bm\theta\in\mathbb T^N$, coupled through
\begin{equation}
\bm q=\bm x-\mathsf L\bm\theta,
\qquad U(\bm x,\bm\theta)=V(\bm q),
\label{eq:matrix_q}
\end{equation}
where $\mathsf L$ is an $m\times N$ gearing matrix compatible with the torus
periods. Let $\mathsf M_x$ and $\mathsf M_\theta$ be constant symmetric
positive-definite mobility matrices. The overdamped dynamics is
\begin{align}
\dot{\bm x}&=\mathsf M_x[-\bm\nabla V(\bm q)-\bm f]
+\sqrt{2T\mathsf M_x}\,\bm\xi_x,
\label{eq:matrix_x}\\
\dot{\bm\theta}&=\mathsf M_\theta[\bm\tau+
\mathsf L^{\mathsf T}\bm\nabla V(\bm q)]
+\sqrt{2T\mathsf M_\theta}\,\bm\xi_\theta.
\label{eq:matrix_theta}
\end{align}
Under the same ergodicity and translation-invariance assumptions as in
Sec.~\ref{sec:reduction}, the stationary density depends only on $\bm q$.
Writing
$\bm g(\bm q)=\bm\nabla V+T\bm\nabla_{q}\ln\rho$, the local velocities are
$\bm\nu_x=-\mathsf M_x(\bm g+\bm f)$ and
$\bm\nu_\theta=\mathsf M_\theta(\bm\tau+
\mathsf L^{\mathsf T}\bm g)$. Eliminating $\bm g$ yields the matrix
reciprocal-current identity
\begin{equation}
\boxed{
\bm\nu_\theta(\bm q)=\mathsf M_\theta
(\bm\tau-\mathsf L^{\mathsf T}\bm f)
-\mathsf M_\theta\mathsf L^{\mathsf T}
\mathsf M_x^{-1}\bm\nu_x(\bm q).}
\label{eq:matrix_identity}
\end{equation}
At vector stall, $\bm J_x=\avg{\bm\nu_x}=\bm0$, and therefore
\begin{equation}
\bm J_\theta^{\st}=\mathsf M_\theta
(\bm\tau-\mathsf L^{\mathsf T}\bm f_{\st}).
\label{eq:matrix_Jtheta}
\end{equation}
Define
\begin{equation}
\mathsf B=\mathsf L\mathsf M_\theta\mathsf L^{\mathsf T},
\qquad
\mathsf G=\mathsf M_x^{-1}
+\mathsf M_x^{-1}\mathsf B\mathsf M_x^{-1},
\label{eq:BG}
\end{equation}
and the reciprocal observed functional
\begin{equation}
\Psi_x^{\st}=\frac{1}{T}
\avg{(\bm\nu_x^{\st})^{\mathsf T}\mathsf G\bm\nu_x^{\st}}.
\label{eq:Psi_matrix}
\end{equation}
The cross term generated by Eq.~(\ref{eq:matrix_identity}) vanishes at vector
stall, giving
\begin{equation}
\ep^{\st}=\Psi_x^{\st}
+\frac{1}{T}(\bm J_\theta^{\st})^{\mathsf T}
\mathsf M_\theta^{-1}\bm J_\theta^{\st}.
\label{eq:matrix_ep}
\end{equation}
The steady power balance is
$T\ep=\bm\tau^{\mathsf T}\bm J_\theta-
\bm f^{\mathsf T}\bm J_x$. Combining it with
Eq.~(\ref{eq:matrix_Jtheta}) and (\ref{eq:matrix_ep}) gives the scalar matrix
closure
\begin{equation}
\boxed{
\bm f_{\st}^{\mathsf T}\bm J_\parallel^{\st}=T\Psi_x^{\st},
\qquad
\bm J_\parallel=\mathsf L\bm J_\theta .}
\label{eq:matrix_closure}
\end{equation}
Unlike the one-dimensional case, one scalar power identity does not determine
the full vector $\bm J_\parallel$; it fixes only its component conjugate to the
stall load.

The hidden-current geometry is nevertheless exact. With $\mathsf B^+$ the
Moore--Penrose pseudoinverse, define
\begin{equation}
\bm J_\theta^{\parallel}=\mathsf M_\theta\mathsf L^{\mathsf T}
\mathsf B^+\bm J_\parallel,
\qquad
\bm J_\theta^{\perp}=\bm J_\theta-\bm J_\theta^{\parallel}.
\label{eq:matrix_projection}
\end{equation}
Then $\mathsf L\bm J_\theta^{\perp}=\bm0$ and
\begin{equation}
\bm J_\theta^{\mathsf T}\mathsf M_\theta^{-1}\bm J_\theta
=\bm J_\parallel^{\mathsf T}\mathsf B^+\bm J_\parallel
+(\bm J_\theta^{\perp})^{\mathsf T}
\mathsf M_\theta^{-1}\bm J_\theta^{\perp}.
\label{eq:matrix_pythag}
\end{equation}
Consequently,
\begin{equation}
\ep^{\st}=\Psi_x^{\st}
+\frac{1}{T}(\bm J_\parallel^{\st})^{\mathsf T}
\mathsf B^+\bm J_\parallel^{\st}
+\mathcal D_\perp^{\st},
\label{eq:matrix_decomp}
\end{equation}
where
$\mathcal D_\perp^{\st}=T^{-1}(\bm J_\theta^{\perp,\st})^{\mathsf T}
\mathsf M_\theta^{-1}\bm J_\theta^{\perp,\st}\ge0$.
The generalized Cauchy--Schwarz inequality and
Eq.~(\ref{eq:matrix_closure}) give
\begin{equation}
\boxed{
\ep^{\st}\ge\Psi_x^{\st}
+\frac{T(\Psi_x^{\st})^2}
{\bm f_{\st}^{\mathsf T}\mathsf B\bm f_{\st}}}
\label{eq:matrix_bound}
\end{equation}
when $\bm f_{\st}^{\mathsf T}\mathsf B\bm f_{\st}>0$. Equality requires both
$\bm J_\theta^{\perp}=0$ and
$\bm J_\parallel\propto\mathsf B\bm f_{\st}$. For $m=1$,
$\Psi_x=K\Phi_x$ and Eq.~(\ref{eq:matrix_bound}) reduces exactly to
Eq.~(\ref{eq:bound_K}). The number of hidden mean-current directions that can
remain completely invisible is
\begin{equation}
\dim\ker\mathsf L=N-\operatorname{rank}\mathsf L.
\label{eq:nullity}
\end{equation}
Thus each independent observed interaction coordinate removes at most one
hidden direction from the no-go sector.

\subsection{Configuration-dependent mobility: the covariance correction}

The pointwise elimination does not require constant mobility. Suppose the
thermodynamically consistent Fokker--Planck currents have the form
\begin{align}
\bm j_x&=\mathsf M_x(\bm q)[-\bm\nabla V-\bm f]p
-T\mathsf M_x(\bm q)\bm\nabla_xp,\\
\bm j_\theta&=\mathsf M_\theta(\bm q)
[\bm\tau+\mathsf L^{\mathsf T}\bm\nabla V]p
-T\mathsf M_\theta(\bm q)\bm\nabla_\theta p,
\end{align}
with positive-definite matrices at every $\bm q$. The local relation becomes
\begin{equation}
\bm\nu_\theta=\mathsf M_\theta(\bm q)
(\bm\tau-\mathsf L^{\mathsf T}\bm f)
-\mathsf M_\theta(\bm q)\mathsf L^{\mathsf T}
\mathsf M_x(\bm q)^{-1}\bm\nu_x.
\label{eq:variable_identity}
\end{equation}
Let $\bm d=\bm\tau-\mathsf L^{\mathsf T}\bm f_{\st}$ and
\begin{equation}
\mathsf G(\bm q)=\mathsf M_x(\bm q)^{-1}
+\mathsf M_x(\bm q)^{-1}\mathsf L\mathsf M_\theta(\bm q)
\mathsf L^{\mathsf T}\mathsf M_x(\bm q)^{-1}.
\end{equation}
Expanding the current-square entropy production gives the exact stall identity
\begin{align}
\ep^{\st}={}&\frac{1}{T}\avg{
(\bm\nu_x^{\st})^{\mathsf T}\mathsf G(\bm q)\bm\nu_x^{\st}}
+\frac{1}{T}\avg{\bm d^{\mathsf T}\mathsf M_\theta(\bm q)\bm d}
\notag\\
&-\frac{2}{T}\avg{
\bm d^{\mathsf T}\mathsf M_\theta(\bm q)\mathsf L^{\mathsf T}
\mathsf M_x(\bm q)^{-1}\bm\nu_x^{\st}}.
\label{eq:variable_ep}
\end{align}
The last term is a mobility--current covariance. It is not removed by
$\avg{\bm\nu_x}=0$ because its coefficient depends on configuration. Thus the
constant-mobility closure and the equality
$\dot Q_x/T=\Phi_x$ do not extend unchanged: either the covariance must be
measured or additional structure must make its coefficient constant. Equation
(\ref{eq:variable_ep}) identifies precisely the extra information required and
marks the boundary of the simple reciprocal reconstruction.

\section{Direct numerical test of a reciprocally geared motor}
\label{sec:numerics}

We test the analytical identities using two hidden phases and a sinusoidal
interaction,
\begin{equation}
L=2\pi,
\qquad V(q)=V_0(1-\cos q),
\label{eq:num_potential}
\end{equation}
with
\begin{equation}
T=\mu_x=1,
\quad \mu_1=\mu_2=\frac12,
\quad \ell_1=\ell_2=1,
\quad V_0=2.5,
\label{eq:num_params}
\end{equation}
and opposing redistributions of a fixed total hidden drive,
\begin{equation}
\tau_1=\tau_0+\delta,
\qquad \tau_2=\tau_0-\delta,
\qquad \tau_0=1.2.
\label{eq:num_torques}
\end{equation}
The reduced $q$ dynamics and every observed statistic are independent of
$\delta$, because they depend on
$\mu_1\tau_1+\mu_2\tau_2=\tau_0$. Numerical evaluation of the exact quadrature
(\ref{eq:Jq}) gives
\begin{equation}
\begin{aligned}
f_{\st}&=0.80548, & \Phi_x^{\st}&=0.15889,\\
J_\parallel^{\st}&=\tau_0-f_{\st}=0.39452.&&
\end{aligned}
\label{eq:num_values}
\end{equation}
Here $K=2$ and $B=1$. At stall,
\begin{equation}
J_1^{\st}=\frac12(\tau_0+\delta-f_{\st}),
\qquad
J_2^{\st}=\frac12(\tau_0-\delta-f_{\st}),
\end{equation}
so the orthogonal dissipation is exactly
$\mathcal D_\perp^{\st}=\delta^2/T$. The total result is therefore
\begin{equation}
\boxed{
\ep^{\st}(\delta)=
2\Phi_x^{\st}+(\tau_0-f_{\st})^2+\delta^2
=0.47342+\delta^2.}
\label{eq:num_ep_delta}
\end{equation}
The first two terms are completely fixed by the observed reciprocal channel;
the last can be varied arbitrarily without changing the observed process.

\begin{figure*}[t]
\includegraphics[width=0.98\textwidth]{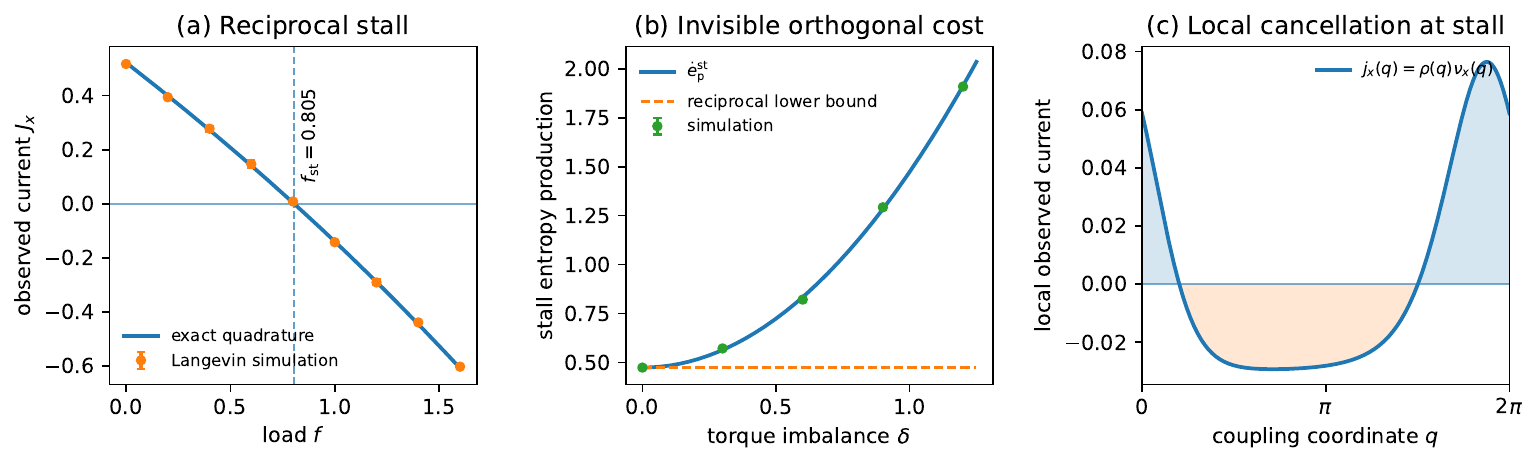}
\caption{Direct test of the reciprocal theory for
Eqs.~(\ref{eq:num_potential})--(\ref{eq:num_torques}).
(a) Exact observed current from the tilted-periodic quadrature and simulation
means; the dashed line marks $f_{\st}=0.80548$.
(b) Total stall entropy production versus the torque imbalance $\delta$.
The exact curve is $0.47342+\delta^2$ and the dashed line is the fixed reciprocal
lower bound. The observed process is identical for every $\delta$.
(c) Exact local observed current $j_x(q)=\rho(q)\nu_x(q)$ at stall. Its signed
integral is zero while its square contributes positively to $\Phi_x^{\st}$.
Simulation points are means of 10--12 independent trajectories with time step
$10^{-3}$, burn-in times 150--200, and production times 1350--1800; error bars
are standard errors of the mean and are smaller than some markers.}
\label{fig:geared_validation}
\end{figure*}

Figure~\ref{fig:geared_validation} compares exact quadrature with direct
Euler--Maruyama simulations of the full three-coordinate Langevin equations.
Panel (a) verifies the current--load relation and the predicted stall load.
Panel (b) keeps the entire observed process fixed while increasing $\delta$;
the measured entropy production follows Eq.~(\ref{eq:num_ep_delta}), whereas
the reciprocal lower bound remains constant. Panel (c) shows the $q$-resolved
observed probability current at stall. Its integral vanishes, but positive and
negative local pieces survive, giving $\Phi_x^{\st}>0$.

The numerical experiment also supplies a stringent identifiability test. The
load curve, stationary density of $q$, $x$-trajectory statistics, stall force,
and $\Phi_x^{\st}$ are identical for all $\delta$. Only measurements that
resolve an additional hidden direction can distinguish the motors. This is the
finite-parameter realization of the null-space statement
Eq.~(\ref{eq:nullity}).

\section{Discussion}
\label{sec:discussion}

The results establish three distinct levels of observation. At the weakest
level, the stationary one-point marginal and mean displacement current are
silent at stall. They cannot distinguish equilibrium from local current
cancellation. At the passive dynamical level, the time-reversal asymmetry of
$x(t)$ gives the rigorous bound~(\ref{eq:path_bound}) and can reveal
irreversibility without a mean current, but it may miss hidden cycles in the
null space of the coupling. At the active dynamical level, a calibrated response
measurement is stronger: for the constant-mobility rank-one model,
Eqs.~(\ref{eq:Qx_equals_phi}) and (\ref{eq:Phi_HS}) reconstruct the exact
reciprocal functional $\Phi_x$ from $x(t)$ alone. The stall load then closes the
projected hidden current and yields the lower bound on total dissipation.

The physical mechanism remains simple. Stall balances only the average force
transmitted to the observed coordinate. Internal-state dependence produces
positive local observed currents in some configurations and negative currents
in others. Their average is zero but their current square is not. A second and
logically independent mechanism is hidden circulation orthogonal to the
coupling. Such circulation neither changes the reduced coordinate nor the
entire observed path measure. Reciprocal coupling makes the first mechanism
inferable; no measurement of the single observed channel can determine the
second without additional structural assumptions or observables.

The matrix formulation makes this geometry explicit. The range of
$\mathsf L$ is the hidden-current sector that can affect the observed
interaction coordinates, while $\ker\mathsf L$ is thermodynamically active but
mechanically invisible. The weighted pseudoinverse in
Eq.~(\ref{eq:matrix_projection}) is the least-dissipative hidden realization of
a given observed projection, and Eq.~(\ref{eq:matrix_bound}) is the associated
sharp projection bound. The simulation in Fig.~\ref{fig:geared_validation}
shows this geometry directly: varying $\delta$ moves only in the null space,
leaving every observed statistic fixed while adding exactly $\delta^2$ to the
stall dissipation.

The nonreciprocal clock and the two-reservoir bookkeeping delimit the role of
mechanical structure. Without back-reaction, the observed current--load curve
places no exact constraint on the hidden clock dissipation. In a heat engine,
calorimetry can still expose a hidden leak, and
Eq.~(\ref{eq:eta_carnot}) shows that every positive total entropy production
reduces the efficiency below Carnot. Mechanical stall, zero observed power, and
zero observed mean current are therefore much weaker than reversibility in both
the isothermal and two-temperature descriptions.

Several limitations remain. The exact fluctuation--response reconstruction uses
constant friction, additive thermal noise, stationarity, and a single
rank-one reduced coordinate. Configuration-dependent mobilities generate the
covariance in Eq.~(\ref{eq:variable_ep}); underdamped dynamics requires velocity
parity and the corresponding inertial Harada--Sasa formulation; multiple
reservoirs require reservoir-resolved currents. Spatially varying temperature
and time-dependent driving can be treated within broader entropy-balance
frameworks
\cite{Taye2020EntropyProductionEntropy,Taye2025ThermodynamicIrreversibilityUnderdamped,Duki2018StochasticResonanceFirst2,Birhanu2021StochasticResonatorLayered,Asfaw2013EffectThermalInhomogeneity,Taye2023TimeDependentSolutions,Taye2025DirectedTransportShort,Taye2015RectifiedMotionShort,Taye2026BrownianMotorsBrownian},
but the observable reconstruction must then be rederived. Experimentally, the
most direct next test is a geared colloidal or molecular system in which a probe
force can be applied to the tracked coordinate while one additional internal
marker is recorded. Such a measurement would compare the response-based
reciprocal estimate with the reduction of the null-space ambiguity produced by
the extra marker.

\section{Conclusion}
\label{sec:conclusion}

We have shown that a stalled microscopic motor can remain dissipative because
zero observed current is only a projected scalar condition. For a reciprocal
motor with one observed coordinate and many hidden phases, a pointwise current
identity gives the exact stall entropy production, a sharp decomposition into
reciprocal and orthogonal hidden costs, and a lower bound on the total cost. The
static one-point displacement statistics are uninformative at stall, but the
dynamical limitation is weaker than previously implied: the observed-bath heat
is exactly $T\Phi_x$, so the Harada--Sasa correlation--response violation of
$x(t)$ reconstructs the full reciprocal contribution without resolving the
hidden phases. A passive time-reversal estimator remains a universal lower
bound but can miss mechanically invisible loops. The matrix extension shows
that the irreducible hidden sector is precisely the null space of the gearing
matrix, while configuration-dependent mobilities add an explicit covariance
correction. Full Langevin simulations verify the predicted stall load, local
current cancellation, and a tunable orthogonal dissipation that leaves all
observed dynamics unchanged. Together with the nonreciprocal and heat-engine
comparisons, these results sharpen the central distinction: mechanical silence
is not thermodynamic silence, and the amount that can be inferred is fixed by
the geometry of coupling and by whether observation is static, passive, or
actively response-resolved.

\appendix

\section{Entropy balance}
\label{app:entropy}

Starting from $S=-\int p\ln p$ and using $\int\partial_t p=0$,
$\dd S/\dd t=-\int(\partial_t p)\ln p$. Substituting the Fokker--Planck
equation~(\ref{eq:fp}) and integrating by parts on the torus gives
$\dd S/\dd t=-\int(j_x\partial_x\ln p+\sum_\alpha j_\alpha\partial_{\theta_\alpha}\ln p)$.
Writing the currents as $j_x=\mu_x F_x^{\mathrm{tot}}p-\mu_x T\partial_x p$ with
$F_x^{\mathrm{tot}}=-V'(q)-f$, and
$j_\alpha=\mu_\alpha F_\alpha^{\mathrm{tot}}p-\mu_\alpha T\partial_{\theta_\alpha}p$
with $F_\alpha^{\mathrm{tot}}=\tau_\alpha+\ell_\alpha V'(q)$, one has
$\partial_x\ln p=F_x^{\mathrm{tot}}/T-j_x/(\mu_x T p)$ and
$\partial_{\theta_\alpha}\ln p=F_\alpha^{\mathrm{tot}}/T-j_\alpha/(\mu_\alpha T p)$.
Substitution separates the positive current-square terms from the drift terms,
giving $\dd S/\dd t=\ep-\dot h_{\mathrm{d}}$ with $\ep$ and $\dot h_{\mathrm{d}}$
as in Eqs.~(\ref{eq:ep_full}) and (\ref{eq:hd}).

\section{Reduction to the coupling coordinate}
\label{app:reduction}

From $\dot q=\dot x-\sum_\alpha\ell_\alpha\dot\theta_\alpha$ and
Eqs.~(\ref{eq:langevin_x})--(\ref{eq:langevin_theta}),
\begin{equation}
\begin{split}
\dot q={}&-\Bigl(\mu_x+\sum_\alpha\ell_\alpha^2\mu_\alpha\Bigr)V'(q)
-\mu_x f-\sum_\alpha\ell_\alpha\mu_\alpha\tau_\alpha\\
&+\sqrt{2\mu_x T}\,\xi_x-\sum_\alpha\ell_\alpha\sqrt{2\mu_\alpha T}\,\xi_\alpha.
\end{split}
\end{equation}
The interaction terms combine into $-MV'(q)$ with $M$ as in
Eq.~(\ref{eq:M_Lambda}), and the independent noises combine into a single white
noise of variance $2MT$, giving Eq.~(\ref{eq:qdot}) with tilt $\Lambda$.

\section{Tilted-periodic current}
\label{app:tilted}

The stationary reduced current is
$\mathcal J_q=-M[V'(q)+\Lambda]\rho-MT\rho'$. With $\psi=(V+\Lambda q)/T$ this is
$\rho'+\psi'\rho=-\mathcal J_q/(MT)$, and multiplication by $e^{\psi}$ gives
$\dd(e^{\psi}\rho)/\dd q=-(\mathcal J_q/MT)e^{\psi}$. Integrating over one
period and using $\rho(q+L)=\rho(q)$ with $\psi(q+L)=\psi(q)+\Lambda L/T$ yields
\begin{equation}
\rho(q)=\frac{\mathcal J_q}{MT\,(1-e^{\Lambda L/T})}\,
e^{-\psi(q)}\int_q^{q+L}e^{\psi(y)}\,\dd y,
\end{equation}
and normalization gives Eq.~(\ref{eq:Jq}).

\section{Orthogonal decomposition}
\label{app:decomp}

Minimizing $\sum_\alpha J_\alpha^2/\mu_\alpha$ subject to
$\sum_\alpha\ell_\alpha J_\alpha=J_\parallel$ with a Lagrange multiplier gives
$J_\alpha/\mu_\alpha=\lambda\ell_\alpha$, hence
$J_\alpha=\lambda\mu_\alpha\ell_\alpha$ and $\lambda=J_\parallel/B$ with
$B=\sum_\alpha\ell_\alpha^2\mu_\alpha$, so
$J_\alpha^{\parallel}=(\mu_\alpha\ell_\alpha/B)J_\parallel$ and the minimum is
$J_\parallel^2/B$. Writing $J_\alpha=J_\alpha^{\parallel}+J_\alpha^{\perp}$ with
$\sum_\alpha\ell_\alpha J_\alpha^{\perp}=0$, the cross term
$\sum_\alpha J_\alpha^{\parallel}J_\alpha^{\perp}/\mu_\alpha
=(J_\parallel/B)\sum_\alpha\ell_\alpha J_\alpha^{\perp}$ vanishes, giving
Eq.~(\ref{eq:pythag}).

\section{Uniform steady state of the solvable model}
\label{app:uniform}

Because the drifts in Eqs.~(\ref{eq:solv_x})--(\ref{eq:solv_theta}) are
independent of $x$, the stationary density factorizes as
$p^{\sstate}=\rho(\theta)/L$. The marginal phase equation
$\Omega\rho-D_\theta\rho'=\mathrm{const}$ with periodic boundary conditions has
only the uniform solution; for $\Omega\neq0$ the exponential solution
$\propto e^{\Omega\theta/D_\theta}$ is not periodic, so $\rho=1/(2\pi)$, and for
$\Omega=0$ periodicity again forces $\rho=1/(2\pi)$. Evaluating
Eq.~(\ref{eq:ep_full}) on $p^{\sstate}=1/(2\pi L)$ with the
currents~(\ref{eq:solv_currents}) gives, for the observed channel,
$\dot e_{\mathrm{p},x}^{\sstate}=(\mu/2\pi T)\int_0^{2\pi}[F(\theta)-f]^2\,\dd\theta$; the
decomposition $F-f=(F-\bar F)+(\bar F-f)$ with vanishing cross term yields
$\dot e_{\mathrm{p},x}^{\sstate}=(\mu/T)[V_F+(\bar F-f)^2]$, and the hidden channel gives
$\dot e_{\mathrm{p},\theta}^{\sstate}=\Omega^2/D_\theta$, so Eq.~(\ref{eq:solv_ep_load}) follows. The
entropy extraction rate $\dot h_{\mathrm{d}}^{\sstate}$ evaluated on the same state
equals $\ep^{\sstate}(f)$ term by term, confirming the steady-state balance.

\section{Matrix projection and proof of the bound}
\label{app:matrix_projection}

For fixed $\bm J_\parallel$ in the range of $\mathsf L$, minimize
$\bm J_\theta^{\mathsf T}\mathsf M_\theta^{-1}\bm J_\theta$ subject to
$\mathsf L\bm J_\theta=\bm J_\parallel$. Introducing a multiplier $\bm\lambda$
gives
$2\mathsf M_\theta^{-1}\bm J_\theta-2\mathsf L^{\mathsf T}\bm\lambda=0$,
so $\bm J_\theta=\mathsf M_\theta\mathsf L^{\mathsf T}\bm\lambda$. The
constraint gives $\mathsf B\bm\lambda=\bm J_\parallel$ and the minimum-norm
solution is $\bm\lambda=\mathsf B^+\bm J_\parallel$, yielding
Eq.~(\ref{eq:matrix_projection}). For any
$\bm J_\theta^{\perp}\in\ker\mathsf L$, the cross term is
\begin{equation}
(\bm J_\theta^{\parallel})^{\mathsf T}\mathsf M_\theta^{-1}
\bm J_\theta^{\perp}
=\bm J_\parallel^{\mathsf T}\mathsf B^+
\mathsf L\bm J_\theta^{\perp}=0,
\end{equation}
which proves Eq.~(\ref{eq:matrix_pythag}). Because
$\bm J_\parallel$ belongs to the range of $\mathsf B$,
\begin{equation}
(\bm f^{\mathsf T}\bm J_\parallel)^2
\le(\bm f^{\mathsf T}\mathsf B\bm f)
(\bm J_\parallel^{\mathsf T}\mathsf B^+\bm J_\parallel).
\end{equation}
Using the closure~(\ref{eq:matrix_closure}) and dropping the non-negative
orthogonal term gives Eq.~(\ref{eq:matrix_bound}).

\section{Numerical quadrature and simulation protocol}
\label{app:numerics}

For the sinusoidal test, Eq.~(\ref{eq:Jq}) was evaluated on a uniform grid over
one period. The inner integral was accumulated on a doubled interval
$[0,2L]$, and the density was reconstructed from
\begin{equation}
\rho(q)=\frac{\mathcal J_q}{MT[1-e^{\Lambda L/T}]}
 e^{-\psi(q)}\int_q^{q+L}e^{\psi(y)}\,\dd y.
\end{equation}
The stall load was obtained by a bracketed root of Eq.~(\ref{eq:stall_condition}).
The full Langevin system, rather than only the reduced coordinate, was integrated
with Euler--Maruyama. The plotted load points use ten independent trajectories,
a time step $10^{-3}$, burn-in time 150, and production time 1350. The
orthogonal-dissipation points use twelve trajectories, burn-in time 200, and
production time 1800. Currents were estimated from unwrapped coordinate
displacements. The entropy production was obtained from the steady power
balance
\begin{equation}
T\ep=\tau_1J_1+\tau_2J_2-fJ_x,
\end{equation}
which avoids numerical differentiation of noisy trajectories. The error bars in
Fig.~\ref{fig:geared_validation} are standard errors across independent
realizations. Source code and the numerical data accompanying the manuscript
reproduce the figure without fitted parameters.

\section{Finite-bandwidth reconstruction and error propagation}
\label{app:bandwidth}

For completeness, write the measured correlation and response after a common
linear detector filter $H(\omega)$ as
$\widetilde C_{vv}^{\mathrm m}=|H|^2\widetilde C_{vv}$ and
$\widetilde R_v^{\mathrm m}=H\widetilde R_v$ when the reported velocity is the
filtered output and the probe is defined at the input. A direct substitution
into Eq.~(\ref{eq:Phi_HS}) is valid only after accounting for this transfer
function; otherwise detector attenuation can mimic a violation of the
fluctuation--response relation. Equivalently, one may apply the same digital
filter to the input probe and the velocity record and formulate both spectra in
the filtered variables. The essential requirement is that correlation and
response refer to the same observable.

Let $\widehat\Phi_x$ be obtained by numerical integration over frequency bins
and let $\Sigma$ be the estimated covariance matrix of the binned
correlation--response differences. For weights $w_k$ implementing the chosen
quadrature, the statistical variance is
\begin{equation}
\operatorname{Var}(\widehat\Phi_x)=
\frac{1}{(\mu_xT)^2}\sum_{k,l}w_k\Sigma_{kl}w_l.
\label{eq:Phi_variance}
\end{equation}
Block resampling in the time domain or independent repeated trajectories can be
used to estimate $\Sigma$ without assuming independent Fourier bins. The stall
load and calibration parameters should be propagated jointly because
Eqs.~(\ref{eq:Jpar_cutoff}) and (\ref{eq:confidence_bound}) contain the ratios
$\Phi_x/f_{\st}$ and $\Phi_x^2/f_{\st}^2$. Close to a zero stall load these
ratios become ill-conditioned, consistently with the analytical result that the
reciprocal closure collapses at $f_{\st}=0$. In that regime one should report
$\Phi_x$ itself and avoid inferring $J_\parallel$ by division.

\end{document}